ARTICLE

# Performance vs Portability in Heterogeneous HPC Environments: Why Pre-execution Benchmarking is Required†

Mindaugas Macernis *[a]

Cloud computing and high-performance computing (HPC) typically follow different paradigms: cloud services are often orchestrated using Kubernetes, whereas HPC workloads are managed through batch schedulers such as Slurm. Growing demand for shared computational resources increases the need for interoperability between these environments. This study uses Podman as a user-accessible tool to benchmark quantum chemistry software distributed as precompiled executables. Containerized and on-system execution are compared across CPU-optimized builds and multiple compute nodes, without additional network optimizations. The results demonstrate that performance depends on software compilation, startup overhead, and execution configuration. Container preparation can dominate short-task workflows, while long-running calculations require monitoring of computational progress. These findings highlight the importance of workload-specific benchmarking and user-level monitoring when deploying scientific applications in containers on HPC systems.

## Introduction

Cloud computing and high-performance computing (HPC) typically follow different paradigms: the former emphasizes flexible use of shared resources, while the latter prioritizes computational performance, often using specialized, tightly integrated hardware and optimized software stacks.[1, 2] However, the growing use of GPUs for artificial intelligence (AI), particularly large language models (LLMs), is increasing the need for interoperability between these two paradigms.[3-5]

Cloud services are often managed using *Kubernetes*, which orchestrates containerized applications such as websites, APIs, microservices, and data processing services. It allocates resources and maintains the desired application state, including availability and the number of running replicas. *Kubernetes* communicates with container runtimes such as *containerd* or CRI-O through the *Container Runtime Interface* (CRI). These runtimes use *Open Container Initiative* (OCI) standards for container images and execution[6-8].

HPC and scientific computing clusters commonly use workload managers such as *Slurm[9], OpenPBS[10]*, *PBS Professional[11]*, or *TORQUE[12]*. Their primary role is to allocate resources to specific computational jobs and release those resources when the jobs finish. Parallel jobs may use multiple CPU cores across several compute nodes connected by high-speed networks. Communication and coordination are handled by MPI or other systems, such as Linda. Containerized applications can be launched within these jobs using *Apptainer*, *SingularityCE*, or *Podman*. All three support OCI images, although *Apptainer* and *SingularityCE* commonly use the SIF format for execution.

Podman is a standard container tool available from official Red Hat Enterprise Linux repositories and covered by Red Hat training. This enables researchers to prepare OCI images and test and debug containerized applications on their own hardware before deploying them in an HPC environment.

HPC environments focus on computational performance and parallel scalability. Performance is evaluated using standardized benchmarks, such as High Performance Linpack (HPL), which is used for the TOP500 ranking[13], and local tests that assess performance and help identify hardware issues. The same principle of systematic benchmarking also applies to scientific software, although HPL results do not directly predict application performance.

Here, an OCI container image prepared for the selected quantum chemistry software package is tested to develop recommendations for users and identify the limitations of running OCI containers in an HPC environment with a batch scheduling system, a high-speed network, and a parallel file system. The results highlight the importance of benchmarking before deployment and the need for more extensive monitoring than for local runs.

## Results and discussion

### Importance of resource allocations

As shown in the Figure 1, enabling *Slurm's* "*exclusive*" option yields performance comparable to direct local execution without a batch scheduler. Without this option, execution time increases by a factor of 1.5. Exclusive allocation reserves the node's CPUs for the job, while CPU access restrictions are enforced through *cpusets[14]* when *Slurm's cgroup*-based confinement is enabled.

The calculations targeted 32 physical cores, as the software performed best with one computational thread per physical core. One possible explanation for the slowdown is that the non-exclusive allocation included sibling Hyper-Threading (HT) threads on the same physical cores rather than threads on 32 distinct cores; however, this has not been verified. Linux logical CPU numbering alone does not identify sibling threads; explicit CPU topology information is required. Whether two independent applications can efficiently share a physical core through its two HT threads remains a separate question for future investigation. However, such optimization can be difficult to achieve across large heterogeneous systems; therefore, this study focuses on the default Linux scheduling behaviour.

[a] *Institute of Chemical Physics, Faculty of Physics, Vilnius University, Saulėtekio av. 3, Vilnius, LT-10257, Lithuania.*
** Correspondence: mindaugas.macernis@ff.vu.lt; Tel.: +370 5 223 4659*

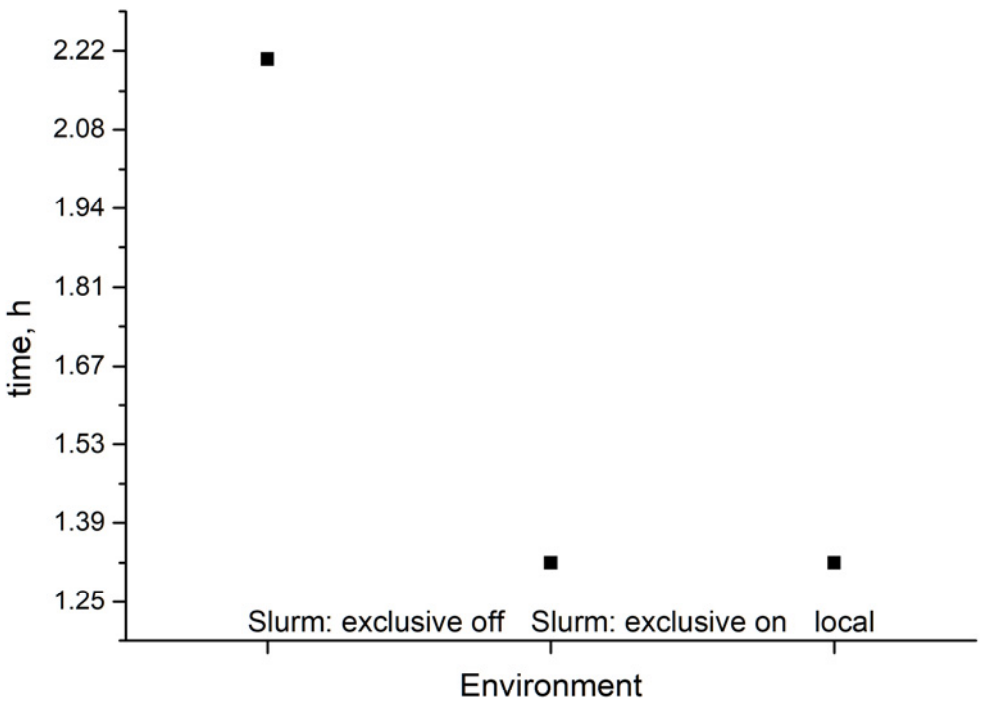

Figure 1. Effect of the computational environment on speedup: local computational performance can be improved by adjusting the “exclusive” option

The next benchmark compares CPU instruction-set optimizations and network interconnects. Figure 2 presents 2 network configurations – *Ethernet* and *InfiniBand* (IB) – and 3 software builds: a legacy x86-64 build and builds optimized for SSE4-enabled and AVX2-enabled, as specified in the technical details.

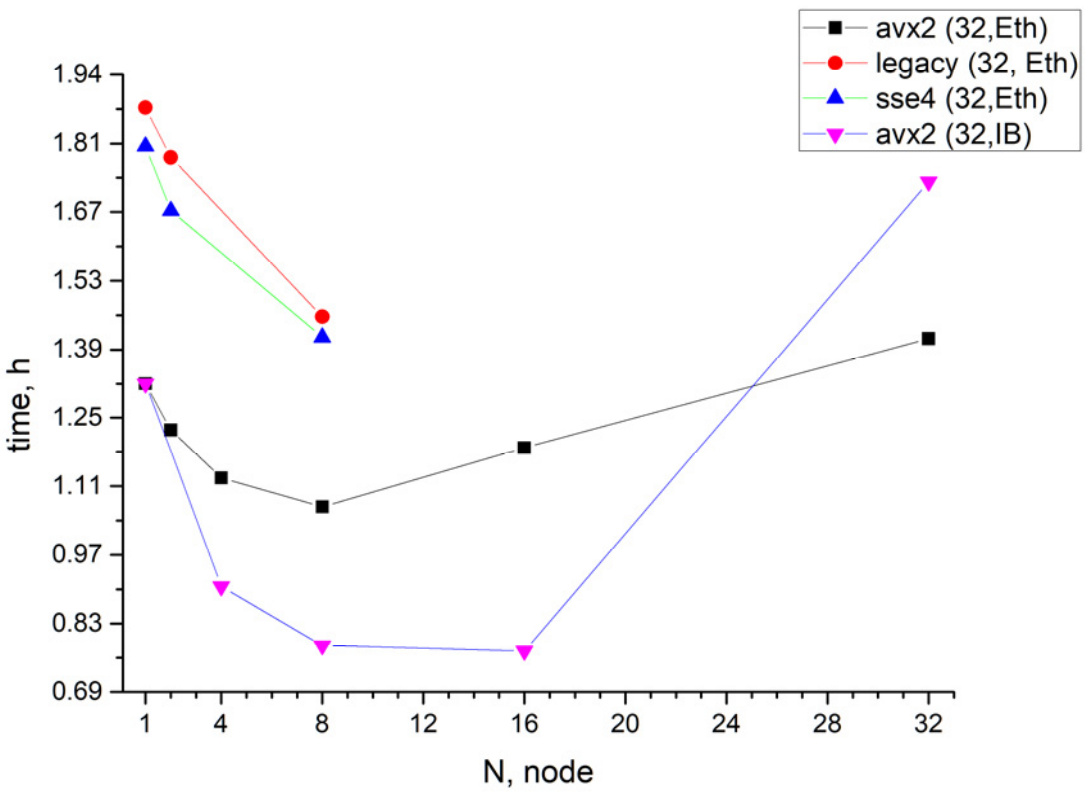

Figure 2. Computational speed-up by using different CPU instructions and network: the network speed-up itself slow down computations despite CPU instruction optimizations.

The results show performance improvements of up to 30% from CPU instruction-set optimizations, with a further improvement of approximately 30% between the two network configurations. Together, these improvements can approach a twofold speedup. These results are illustrative, and their magnitude may depend on the computational workload and problem size.

However, the benchmarks also show a slowdown when scaling beyond 8 nodes with the tested network configuration. This observation should be treated as a system-specific scaling limitation rather than a general rule; it does not by itself establish that the network was misconfigured. Tests using a slower 40 Gbit/s *InfiniBand* interconnect produced different results, with the scaling limit reached at 4 nodes.

Variations in computational performance were also observed when jobs were submitted through *WebMO* (version 20.0.012e)[15], which prepares the execution environment and generates *Slurm* jobs on behalf of users. To limit the number of configuration variables, this study focuses on *Slurm* jobs submitted directly through batch scripts. The influence of additional job submission interfaces and their environment settings requires separate investigation in future work.

**Different size jobs benchmarking**

A parallelizable single-point calculation requiring approximately 5 minutes on a single node was selected to benchmark container performance for small jobs. Table 1 summarizes the results. All jobs were launched from the same compute node, with additional nodes used for multi-node runs. Communication used TCP/IP over *InfiniBand*, with SSH for remote process launch and no additional optimizations.

The CPU time (s) column reports processor usage recorded by the computational software, providing a metric for internal code profiling and debugging rather than elapsed-time analysis. The single-node “On system” run serves as the reference, with a software-reported CPU time of 10,794.9 s, a wall-clock time of 348.6 s (Computing Time), and a total execution time of 351 s. The difference of approximately 2.4 s represents overhead outside the software-reported interval, including software start-up. These timings exclude batch-workflow preparation and clean-up procedures, such as compute-node preparation and scratch-directory clean-up.

Table 1. Benchmarking small computational tasks with Podman: image preparation accounts for a substantial portion of the total execution time. The legal (4.4 GB) and AVX2-enabled (17 GB) images differ in size by a factor of 3.86.

| Instruction Sets | Environment | Nodes | CPU(s) | Computing Time (s) | Total (s) |
|---|---|---|---|---|---|
| legal | Container | 1 | 12194.2 | 392.0 | 472 |
| avx2-enabled | Container | 1 | 10995.0 | 354.8 | 727 |
| | | 2 | 60.9 | 308.4 | 691 |
| | | 4 | 60.9 | 217.2 | 609 |
| | | 8 | 59.8 | 175.3 | 580 |
| | On system | 1 | 10794.9 | 348.6 | 351 |
| | | 2 | 63.4 | 289.0 | 295 |
| | | 4 | 59.4 | 220.6 | 227 |
| | | 8 | 59.7 | 174.9 | 181 |

For the on-system installation, software start-up took approximately 3-6 s. This overhead could potentially be reduced through system-level optimizations, such as placing the software on local disks. Container start-up showed a strong dependence on image size: the smaller image required 80 s, whereas the AVX2-enabled image required approximately 6-7 min, regardless of the number of nodes used in the tests.

The containerized legal build took approximately 10% longer to execute but ran on both CPU types, with and without AVX-enabled support. As expected, the AVX2-enabled build failed on nodes whose CPUs lacked AVX2-enabled support.

Computing times differed between the containerized and on-system runs, with containerized execution taking approximately 1-20 s longer. Such short calculations commonly form part of potential energy surface scans, which may involve around 200 jobs. In this scenario, the timing analysis indicates that repeated container preparation and software start-up would be the dominant overhead.

Assuming 3 min of computation and 5 min of start-up overhead per job, 200 sequential jobs would require 10 h of computation and 16 h 40 min of overhead, giving a total runtime of 26 h 40 min. This estimate assumes that the start-up overhead is incurred separately for every job. During this overhead, allocated resources remain unavailable to other jobs, reducing useful computational throughput. The associated energy cost would require separate measurement.

Table 2 presents a different scenario: a long-running job for which software-environment start-up, either on-system or within a container, is expected to contribute only a small fraction of the total runtime. A single-node calculation lasting up to 3 hours was selected as the reference. The on-system run had a software-reported wall time of 2 h 53 s, with approximately one additional minute outside this interval. This difference includes loading internal program components and start-up and clean-up procedures not captured by the software’s own timer.

Table 2. Benchmarking larger computational tasks with *Podman*, including a case of continued computation in a non-terminating loop, classified as unsuccessful following on-site analysis of the output data.

| Instruction Sets | Environment | Nodes | CPU(s) | Computing Time (s) | Total (s) |
|---|---|---|---|---|---|
| legal | Container | 1 | 454146.3 | 14329.7 | 14435 |
| avx2-enabled | Container | 1 | 363574.4 | 11498.9 | 11915 |
| | | 2 | 2985.2 | 10465.3 | 11246 |
| | | 4 | - | - | *35487* |
| | | 8 | 3442.0 | 7614.0 | 8520 |
| | On system | 1 | 328464.3 | 10390.6 | 10420 |
| | | 2 | 2842.1 | 10215.6 | 10568 |
| | | 4 | 3184.9 | 9253.4 | 9634 |
| | | 8 | 2979.8 | 6613.7 | 6988 |

The results include a case in which the containerized calculation continued for more than 9 hours. Such behaviour may involve an alternative computational pathway used by the software to work around slow I/O, rather than repeated I/O recovery attempts. Identifying this pathway requires debugging instrumentation that cannot be enabled during the ongoing calculation. Moreover, restarting the job does not necessarily reproduce the conditions that triggered it, as illustrated by the successful rerun in this case.

These observations highlight the need to monitor computational progress and investigate or rerun tasks whose execution times substantially exceed an established benchmark. An unusually long containerized run should therefore not automatically be attributed to faulty hardware or inadequate hardware preparation.

Repeated runs showed variations in total execution time. For the containerized AVX2-enabled build, the ranges were 3 h 2 min 55 s - 3 h 18 min 35 s on one node, 3 h 7 min 26 s - 3 h 10 min 41 s on two nodes, 2 h 36 min 2 s - 9 h 51 min 27 s on four nodes, and 2 h 7 min 25 s - 2 h 22 min on eight nodes. The 4-node range includes the previously identified unsuccessful run, retained to highlight the anomalous behaviour. The single-node legal build required 4 h 0 min 35 s - 4 h 22 min 39 s and was the slowest configuration among the successful runs.

For the on-system installation, the corresponding ranges were 2 h 53 min 40 s - 3 h 22 min 51 s, 2 h 56 min 8 s - 3 h 5 min 42 s, 2 h 7 min 58 s - 2 h 40 min 34 s, and 1 h 53 min 20 s - 1 h 56 min 28 s for 1, 2, 4, and 8 nodes, respectively.

The single-node legal build therefore required approximately twice the execution time of the fastest 8-node configuration, while the successful containerized AVX2-enabled runs fell between these extremes. These comparisons highlight the importance of the software build and CPU instruction-set optimizations, alongside node count and execution environment.

## Computational methods

### Hardware & Software

The benchmarks were performed on compute nodes with the following configurations:

- Intel Xeon Gold 6130 processors at 2.10 GHz, Hyper-Threading enabled, 376 GB RAM, and 100 Gbit/s *InfiniBand* connectivity.
- Intel Xeon X5650 processors at 2.67 GHz, Hyper-Threading enabled, 96 GB RAM, and 40 Gbit/s *InfiniBand* connectivity.

The *Ethernet* network used in the tests operated at 1 Gbit/s.

The software environment comprised Gaussian 16 with AVX2-enabled build and without it, Revision C.01[16]; Red Hat Enterprise Linux 8.4 (Ootpa); Slurm 20.11.7-Bull.1.2; and Podman 3.2.3.

Figure 3 shows the structure of the fucoxanthin-acetonitrile[17, 18] complex used in the benchmarks. Calculations were performed at the B3LYP/cc-pVDZ level optimization without basis-set transformation optimizations[19]. This system is related to previous molecular dynamics studies, although its potential energy surface (PES) remains insufficiently characterized.[20]

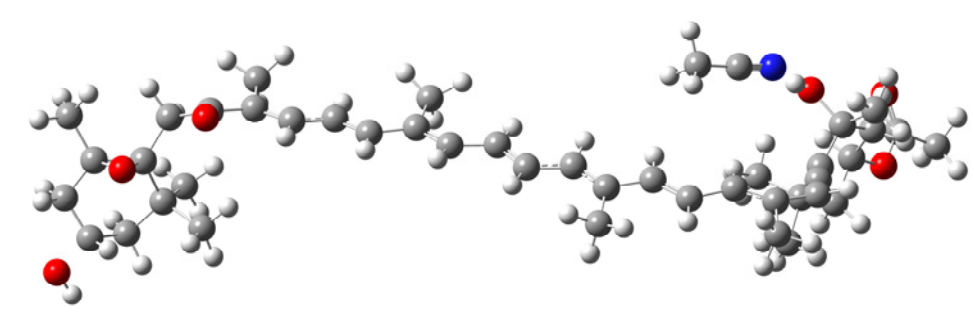

Figure 3. Molecular structure of the fucoxanthin-acetonitrile complex used in the benchmarks.

## Conclusions

These performance benchmarks highlight the need for benchmarking at multiple levels as standardized cloud environments and specialized HPC systems become increasingly interconnected through demand for shared GPU computing resources. In the tested configurations, containerized execution took up to approximately twice as long as the on-system reference. Workflows comprising many short tasks were particularly affected by repeated container preparation and software start-up, which can dominate total execution time. Additional *InfiniBand* and GPU optimizations were outside the scope of this study and require separate evaluation.

Containerized execution can therefore provide a practical starting point for deploying and evaluating scientific applications on HPC systems, or for running long single-node calculations where start-up overhead is relatively small and computational speedup is not a primary requirement. Its suitability for production workflows should be established through workload-specific benchmarks, hardware compatibility checks, and monitoring of computational progress.

## Author contributions

CRediT: **Mindaugas Macernis**: Conceptualization, Data curation, Formal Analysis, Funding acquisition, Investigation, Methodology, Project administration, Resources, Software, Supervision, Validation, Visualization, Writing – review & editing.

## Conflicts of interest

There are no conflicts to declare.

## Data availability

All data supporting the findings of this study are included in the article.

## Acknowledgements

Computations were performed on resources at the supercomputer "VU HPC" Saulėtekis of Vilnius University in Faculty of Physics. Thanks to the Eureka Research Council of Lithuania for financial support (Eureka Grant No. 10-052-P-01-012, SPECTRO Urea Biuret Aqua).